\documentclass[11pt]{article}

\usepackage[margin=1in]{geometry}
\usepackage[T1]{fontenc}
\usepackage[utf8]{inputenc}
\usepackage{lmodern}
\usepackage{amsmath,amssymb}
\usepackage{graphicx}
\usepackage{authblk}
\usepackage[numbers,sort&compress]{natbib}
\usepackage[colorlinks=true,linkcolor=blue,citecolor=blue,urlcolor=blue]{hyperref}

\title{Spatial exceptional points induced by hyperbolicity in an electrostatic spherical shell}

\author[1]{Álvaro Buendía\thanks{alvaro.buendia@inl.int}}
\author[1,2]{Marcelo S. Barreiro}
\author[1,3,4]{Nuno M.R. Peres}

\affil[1]{International Iberian Nanotechnology Laboratory, Av. Mestre Jos\'e Veiga s/n, Braga 4715-330, Portugal}
\affil[2]{University of Lorraine, LCP-A2MC, 1 Bd Arago, F-57070 Metz, France}
\affil[3]{Department of Physics and CFUMUP, University of Minho, Gualtar Campus, Braga 4710-057, Portugal}
\affil[4]{POLIMA, University of Southern Denmark, Moseskovvej 67, 5230 Odense M, Denmark}

\date{}

\begin{document}
\maketitle

\begin{abstract}
Exceptional points (EPs) in optics are usually associated
with gain and loss or non-reciprocity. In this work we
show that a shell with spherical anisotropy supports
"spatial EPs" without either. The radial electrostatic
problem maps onto a damped oscillator in which the dielectric anisotropy $\varepsilon_t/\varepsilon_r$ is the control parameter. At the
EP, the two power-law solutions coalesce spatially and a second,
logarithmic solution emerges. Beyond this threshold, the field switches from monotonic
decay to spatial oscillation, enabling multiple Fabry–Pérot-like resonances. This work reveals how spatial exceptional points modify the optical
response of hyperbolic core-shell particles.
\end{abstract}

\section{Introduction}

Exceptional points (EPs) were first introduced by Kato  \cite{Kato1995} as singularities of non-Hermitian matrices, where not only the eigenvalues coalesce, but also the eigenmodes. Initially  considered  purely mathematical objects, EP singularities were later found to have observable physical consequences \cite{Berry1994}.

Historically, non-Hermitian Hamiltonians, $H\neq H^\dagger$, were regarded as non-physical, as they were considered to break energy conservation. 
However, in 1998, Bender and Boettcher \cite{Bender1998} showed that non-Hermitian Hamiltonians with parity--time ($\mathcal{PT}$) symmetry can possess an entirely real spectrum, which also holds more generally for pseudo-Hermitian systems \cite{Mostafazadeh2002}. Even though these systems possess gain and loss, these are compensated, thus guaranteeing energy conservation. 

This finding stimulated extensive research on non-Hermitian physics and on exceptional points in quantum systems \cite{bender_pt_2019,bergholtz_exceptional_2021}. Moreover, as gain and loss are ubiquitous in classical--wave systems, EPs have been demonstrated across photonics\cite{el-ganainy_non-hermitian_2018, miri_exceptional_2019, bergholtz_exceptional_2021}, acoustics \cite{Shi2016} and mechanical metamaterials \cite{Scheibner2020}. Even the transition between the underdamped and overdamped regimes in such a simple system as the damped harmonic oscillator can be explained through the lens of these non-Hermitian singularities \cite{Fernndez2018, Dolfo2018, Heiss2016}. In optics, exceptional points have been exploited for asymmetric mode switching and conversion \cite{Doppler2016, Roy:22}, enhanced non-reciprocal transmission \cite{Laha2020}, and sensing \cite{Li:18}.

In this Letter, we show that a quasi-electrostatic hyperbolic shell hosts an EP in complete absence of gain, loss, or non-reciprocity in the permittivity. The non-Hermiticity that produces the EP is not a property of the Hamiltonian but rather of the companion matrix describing the equation governing the radial part of the electrostatic potential, with the control parameter being the dielectric anisotropy factor of the shell. Crossing the EP switches the field inside the shell from monotonic decay to spatial oscillation or vice-versa, and the transition is the spatial optical analogue of $\mathcal{PT}$-symmetry breaking in quantum systems or the overdamped to underdamped transition in a harmonic oscillator.

\section{Model and mapping to a damped harmonic oscillator}
\begin{figure*}[htb]
\includegraphics[width=1.0\linewidth]{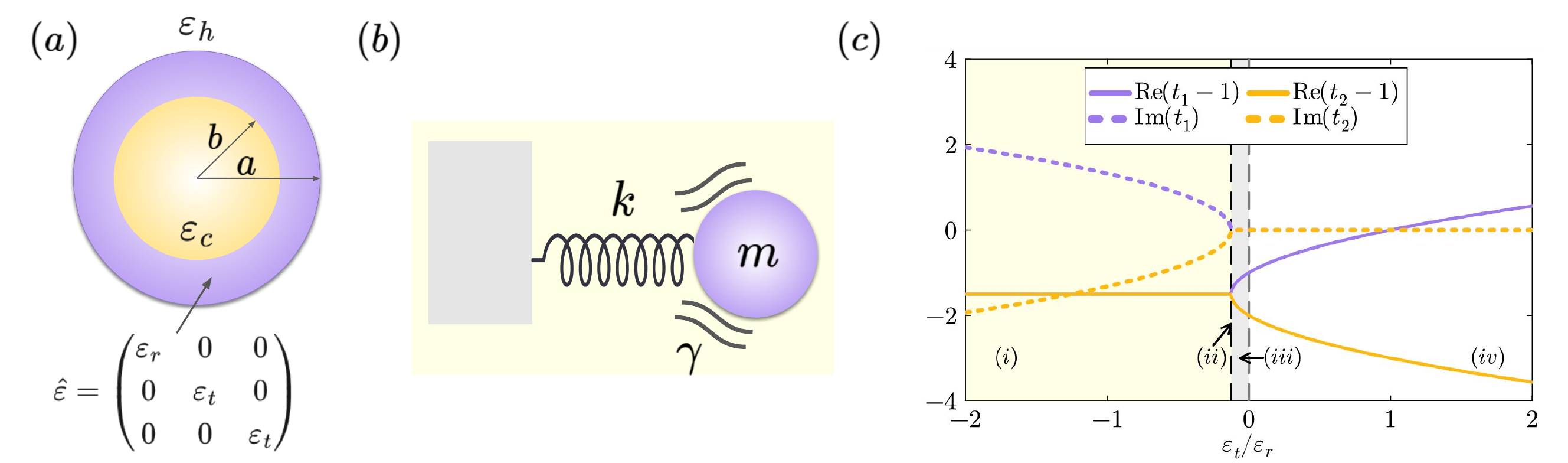}
\caption{(a) Anisotropic uniaxial shell of inner and outer radii $b$ and $a$, respectively, with radial ($\varepsilon_r$) and tangential ($\varepsilon_t$) permittivities. (b) The system can be mapped to a damped harmonic oscillator with stiffness constant $k$, mass $m$ and damping constant $\gamma$. (c) Universal phase diagram: real and imaginary parts of the field exponents $t_{1,2}-1$ as a function of the single control parameter $\varepsilon_t/\varepsilon_r$.From left to right we have the different regimes, (i) In the yellow-shaded region $(i)$, where $\varepsilon_t/\varepsilon_r<-\tfrac18$ the exponents are complex conjugates and the field is oscillatory (underdamped phase).  The exceptional point $(ii)$ sits at $\varepsilon_t/\varepsilon_r=-\tfrac18$ (for $l=1$).  For  $0>\varepsilon_t/\varepsilon_r>-\tfrac18$ (gray-shaded region $(iii)$), the material is in the hyperbolic overdamped phase and the exponents are real (monotonic field). Finally, for $\varepsilon_t/\varepsilon_r > 0$ (non-shaded region) the material is in the elliptic phase $(iv)$. The plot is formally identical to the eigenvalue flow across a $\mathcal{PT}$-symmetry-breaking transition or to the exponents in the damped oscillator.}
\label{fig:phasediagram}
\end{figure*}

In this work, we consider a spherical particle (Fig.~\ref{fig:phasediagram}(a)) with an isotropic core of permittivity $\varepsilon_c$ and radius $b$, and a uniaxial anisotropic shell of outer radius $a$ with permittivity tensor (in spherical coordinates) $\widehat\varepsilon=\operatorname{diag}(\varepsilon_r,\varepsilon_t,\varepsilon_t)$, embedded in a host medium with permittivity $\varepsilon_h$. The problem of scattering in such a system was first solved by Roth and Dignam in 1973 \cite{Roth1973}, as an extension of Mie theory. More recently, particles coated by an anisotropic shell have been studied both in the quasi-static limit \cite{liu_effects_2008, moradi_revisit_2023, Kettunen2013} and beyond \cite{tang_modeling_2018, DeBeule:14}.

Under plane-wave excitation $\mathbf E=E_0\hat{\mathbf x}\,e^{i(k_z z-\omega t)}$ and in the quasi-static regime $ka\ll1$, with the polar axis along $\hat{\mathbf x}$, the electric potential $\phi$ obeys $\nabla\!\cdot(-\hat\varepsilon\nabla\phi)=0$. By separating variables, $\phi(r,\theta) = R(r)\Theta(\theta)$, and writing $s=\log r$ (with $r$ measured in units of the core radius $b$), the radial part becomes the Cauchy--Euler equation  \cite{moradi_revisit_2023}
\begin{equation}
\frac{\partial^{2}R}{\partial s^{2}}+\frac{\partial R}{\partial s}-l(l+1)\,\frac{\varepsilon_t}{\varepsilon_{r}}\,R=0,
\label{eq:cauchy-euler}
\end{equation}
where $l$ is the order of the multipole.

This is the central object of the paper. While this problem has been solved previously, we here recognize that this radial equation has the same functional form as the equation of motion of a damped
harmonic oscillator (see Fig. \ref{fig:phasediagram}(b)),
\begin{equation}
\frac{\partial^2x}{\partial t^2}+\frac{\gamma}{m}\,\frac{\partial x}{\partial t}+\frac{k}{m}\,x=0,
\label{eq:oscillator}
\end{equation}
where $m$ is the mass, $k$ is the stiffness, and $\gamma$ is the damping factor. In the shell, the spatial coordinate $s$ plays the role of time and the
parameters of Eq.~(\ref{eq:cauchy-euler}) map onto
\begin{equation}
   \frac{\gamma}{m}=1,\quad \frac{k}{m}=-\,l(l+1)\frac{\varepsilon_t}{\varepsilon_r},
\end{equation}
where, in the core-shell nanoparticle, the effective damping is fixed by the
spherical geometry, while the effective stiffness
 is tunable and
sign-indefinite. For an elliptic material (either $\varepsilon_t,\varepsilon_r>0$ or $\varepsilon_t,\varepsilon_r<0$), the effective stiffness is negative, while for a hyperbolic material
($\varepsilon_t/\varepsilon_r<0$), the effective stiffness is positive, as in the case of the harmonic oscillator. 

The critical point in the transition between the underdamped and overdamped phases of the oscillator has been identified as an exceptional point \cite{Heiss2016, Dolfo2018, Fernndez2018}. In order to expose the exceptional point in our system, we follow the same procedure and recast \eqref{eq:cauchy-euler}
in first-order form through the companion matrix $M$,
\begin{equation}
\frac{\partial}{\partial s}\begin{pmatrix}R\\
\frac{\partial R}{\partial s}
\end{pmatrix}= M \begin{pmatrix}R\\
\frac{\partial R}{\partial s} \end{pmatrix}, \quad
M=\begin{pmatrix}0&1\\[2pt] l(l+1)\dfrac{\varepsilon_t}{\varepsilon_r}&-1\end{pmatrix},
\label{eq:companion}
\end{equation}
whose eigenvalues $t_i$ and eigenvectors $\boldsymbol{u}_i$ are 
\begin{equation}
t_{1,2}=\tfrac12\!\left(-1\pm\sqrt{1+4\,l(l+1)\tfrac{\varepsilon_t}{\varepsilon_r}}\right),
\quad
\boldsymbol{u}_{1,2}=\begin{pmatrix}1\\ t_{1,2}\end{pmatrix}.
\label{eq:eigen}
\end{equation}
For $\varepsilon_t/\varepsilon_r\neq-1/[4l(l+1)]$, the potential is \cite{liu_effects_2008}
\begin{equation}
\phi(r,\theta)=\begin{cases}
A_1\,r^l P_l(\cos\theta), & r<b,\\[2pt]
(A_2\,r^{t_1}+B_2\,r^{t_2})P_l(\cos\theta), & b<r<a,\\[2pt]
\left(-E_0\,r^l+\dfrac{B_3}{r^{l+1}}\right)P_l(\cos\theta), & r>a,
\end{cases}
\label{eq:potential}
\end{equation}
so that the eigenvalues $t_{1,2}$ of $M$ are the exponents of the potential in the shell, and $t_{1,2}-1$ those of the field.The amplitudes are determined by continuity of $\phi$ and of the normal displacement $\varepsilon_r\partial_r\phi$ at the interfaces $r=a,b$ (see Appendix A). The tangential and radial components of the electric field derive from the potential in Eq.~(\ref{eq:potential}) as $\mathbf E=-\nabla\phi$.  Thus, the field in the shell reads
\begin{equation}
\textbf{E}(r,\theta)
=E_r(r)\, P_l(\cos\theta) \;\hat{r} + E_\theta(r)\, \frac{dP_l(\cos\theta)}{d\theta} \;\hat\theta ,
\label{eq:electricfield}
\end{equation}
where the radial profiles $E_r(r)$ and $E_\theta(r)$ are given by
\begin{equation}
-\begin{pmatrix}E_\theta(r)\\[2pt] E_r(r)\end{pmatrix}
=A_2\,r^{\,t_1-1}\,\boldsymbol{u}_1+B_2\,r^{\,t_2-1}\,\boldsymbol{u}_2,
\label{eq:superposition}
\end{equation}
where $\boldsymbol{u}_i$ are defined in Eq.~(\ref{eq:eigen}) and, for $l=1$, $dP_1(\cos\theta)/d\theta=-\sin\theta$ (see the Appendix A for the full expression in all three regions).

The exponents of the electric field, $t_{1,2}-1$ for $l=1$, are plotted in Fig.~\ref{fig:phasediagram}(c) as a function of $\varepsilon_t/\varepsilon_r$. The solid purple (orange) lines correspond to the real part of $t_1-1$ ($t_2-1$), while the dashed lines represent the imaginary parts. The non-shaded region corresponds to the elliptic phase $\varepsilon_r\varepsilon_t > 0$, whereas the shaded regions represent the hyperbolic phases $\varepsilon_r\varepsilon_t < 0$.

From the damped-oscillator picture, we can immediately organize the physics of the hyperbolic shell into the textbook phases, selected by the sign of the discriminant $1+4l(l+1)\varepsilon_t/\varepsilon_r$ in Eq.~(\ref{eq:eigen}):
\emph{(i)} the hyperbolic shell case with $\varepsilon_t/\varepsilon_r<-1/[4l(l+1)]$ corresponds to the underdamped regime (yellow-shaded region) with complex-conjugate exponents $t_{1,2}=-\tfrac12\pm i\Delta$, with $\Delta =\tfrac12\sqrt{-1-4l(l+1)\,\varepsilon_t/\varepsilon_r}$, and a potential (and thus a field) that oscillates in $s=\log r$ and can be written as
\begin{equation}
\phi_{\rm shell}=r^{-1/2}\!\left[\tilde A_2\cos(\Delta \,\log r)+\tilde B_2\sin(\Delta \,\log r)\right]P_l(\cos\theta),
\end{equation}
where we redefined $\tilde A_2=A_2+B_2$ and $\tilde B_2=i(A_2-B_2)$. 

\emph{(ii)} In the case where $\varepsilon_t/\varepsilon_r=-1/[4l(l+1)]$ the oscillator is critically damped at the EP, where the exponents coalesce at $t=-\tfrac12$. The general solution presented in \cite{moradi_revisit_2023} and previous articles is not valid at the EP, which instead has the Jordan form
\cite{trench2013}
\begin{equation}
\phi_{\rm shell}=r^{-1/2}\!\left(A_2+B_2\log \frac{r}{b
}\right)P_l(\cos\theta).
\end{equation}
The second term, written in terms of $s$, reads $s\,e^{-s/2}$ (a product of a linear and an exponential function), which is the hallmark of critical damping \cite{Fernndez2018} and equivalently, of an exceptional surface wave \cite{Lakhtakia2020,Lakhtakia2020-2}. The explicit form and the coefficients of the electric field at the exceptional point are presented in the Appendix A.

\emph{(iii)}
The hyperbolic (gray-shaded region) shell with  $0>\varepsilon_t/\varepsilon_r>-1/[4l(l+1)]$, corresponding to the overdamped regime, with two real exponents $t_{1,2}$ and a field that decays monotonically. 

\emph{(iv)} The elliptic $\varepsilon_t/\varepsilon_r>0$ case, which corresponds to an oscillator with negative stiffness. One of the exponents becomes positive, which leads to a potential that grows with $r$. Due to this, the two solutions only coexist in a spatially constrained region.

We can see that the emergence of the exceptional point is not limited to the electric dipole; each dipole reaches its own spatial EP at a different anisotropy factor. We only restrict the discussion to $l=1$ for simplicity. We also remark that, although the two distinct phases of the hyperbolic shell have been noted before \cite{moradi_revisit_2023}, their origin as a transition mediated by an
exceptional point has,
to our best knowledge, gone unrecognized. 
\section{Optical response of the anisotropic particle}
In this section we address how this transition affects the optical response. Contrary to the typical exceptional points produced by non-Hermiticity of the Hamiltonian, this spatial exceptional point does not produce a temporal coalescence of optical modes but a spatial one. Nevertheless, the spatial redistribution of the field induced by the exceptional point will modify the optical response of the particle. To illustrate this, we consider a physical realization of the system. 

Recently, plexcitonic particles formed by a metallic core and a J-aggregate layer have been shown experimentally to be well described by uniaxial anisotropic permittivity models \cite{kondorskiy_optical_2024,kondorskiy_manifestation_2022}. Ellipsometry measurements also confirm uniaxial tensor for the permittivity of some thin J-aggregate films \cite{Gama2026}. Depending on the orientation of the resonant elements, the shell can be type-I or type-II hyperbolic within a frequency window (Reststrahlen band). We consider a type-II hyperbolic shell, consistent with tangentially oriented transition dipoles,
\begin{align}
&\varepsilon_r(\omega) = \varepsilon_\infty, \quad
&\varepsilon_t(\omega) = \varepsilon_\infty  + \frac{f \omega_0^2}{\omega_0^2-\omega^2-i \omega\gamma}.
\end{align}

\begin{figure}[!t]
\centering
\includegraphics[width=0.5\linewidth]{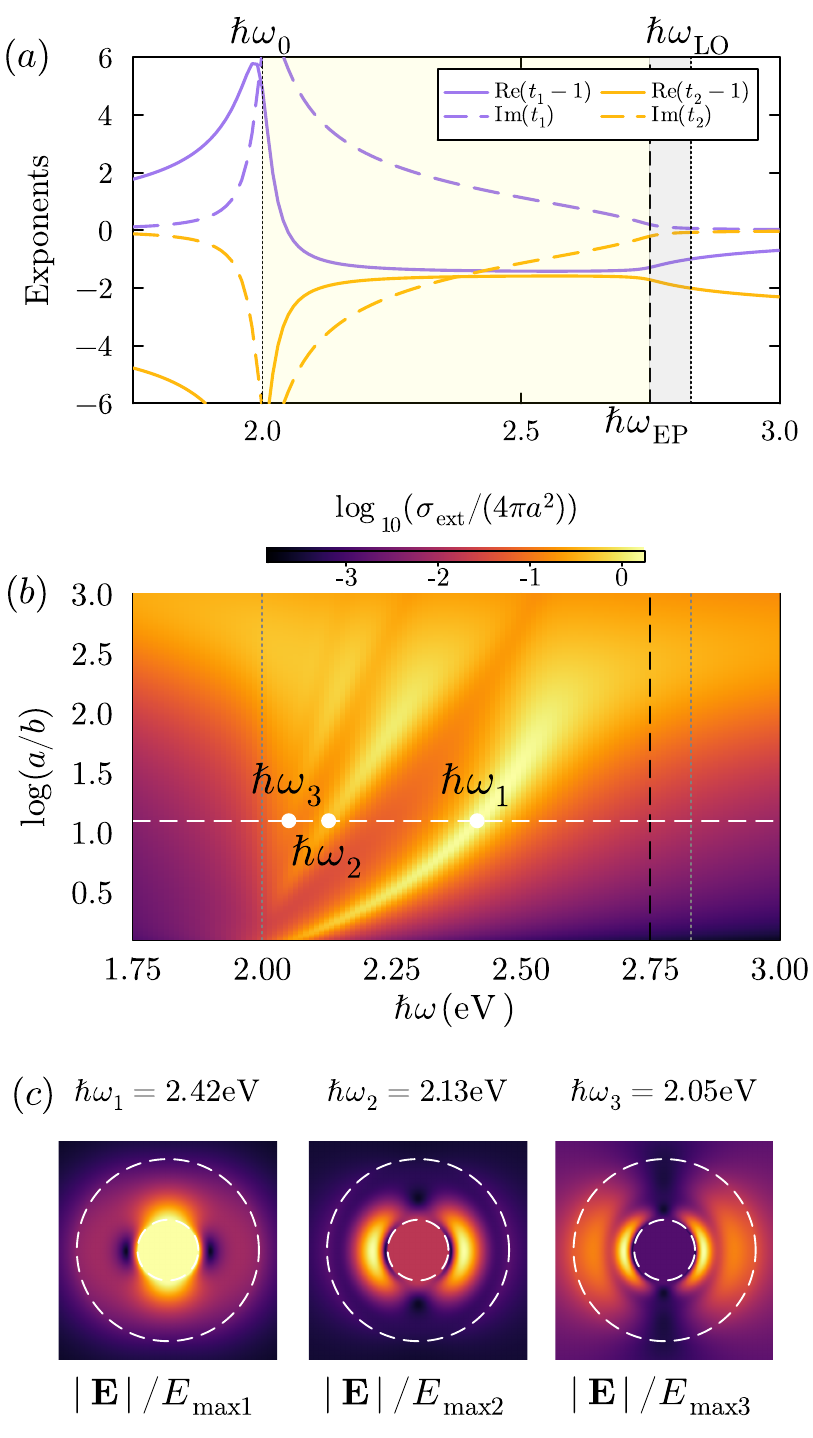}
 \caption{(a) Real (solid) and imaginary (dashed) parts of the field exponents $t_1-1$ (purple) and $t_2-1$ (orange) for the Lorentzian model. The shaded regions mark the Reststrahlen band, where the shell is hyperbolic, with yellow and gray shading indicating the underdamped and overdamped phases. The dashed black line marks the EP, $\hbar\omega_\text{EP}\approx2.75$~eV, and the dotted lines mark the hyperbolic--elliptic transitions at $\hbar\omega_0$ and $\hbar\omega_\text{LO}$. (b) Extinction cross section $\log_{10}({\sigma_\textrm{ext}/(4 \pi a^2)})$ fixing core radius $b=10$~nm and changing $\log(a/b)$ and energy. For $a=30$~nm, $\log(a/b)= \log 3 \approx 1.1$ (dashed white horizontal line) the shell hosts up to $n=3$ resonances, $\hbar\omega_n = 2.42, 2.13, 2.05$~eV (white dots). (c) Normalized field amplitudes $|\mathbf{E}|/E_{\textrm{max}\,n}$ at the resonance frequencies $\hbar\omega_n$ for $b = 10$~nm, with $E_{\textrm{max}\,n}/E_0 = 29.1$, $13.1$, and $4.2$. Dashed circles mark $r=b$ and $r=a$. All resonances lie in the underdamped phase, where the field oscillates within the shell, producing Fabry--P\'erot-like modes.}
\label{fig:Lorentzian}
\end{figure}

This material achieves a Reststrahlen band between $\omega_0$ and $\omega_\textrm{LO} = \sqrt{1+f/\varepsilon_\infty}\, \omega_0$ in the lossless limit. As the degree of anisotropy $\varepsilon_t/\varepsilon_r$ is now $\omega$-dependent, frequency can be a control parameter for the transition. We fix the parameters, $\varepsilon_\infty=\varepsilon_c=\varepsilon_h =1$ and $f=1, \hbar\omega_0 = 2.0\;\textrm{eV}, \hbar\gamma = 0.05\;\textrm{eV}$ and plot in Fig.~\ref{fig:Lorentzian}(a) the real and imaginary parts of the exponents of the dipolar electric field, $t_1 - 1$ and $t_2 - 1$, depending on frequency . For $l=1$ , the nominal EP frequency is $\hbar\omega_\textrm{EP}  =  \hbar\omega_0 \sqrt{(9\varepsilon_\infty +8f)/(9\varepsilon_\infty)} \approx 2.75$~eV. Since $\varepsilon_t/\varepsilon_r$ increases monotonically from $-\infty$ to $0$ across the Reststrahlen band, increasing $\omega$ sweeps the hyperbolic part of Fig.~\ref{fig:phasediagram}(c) from left to right: first the underdamped and then the overdamped phase.

So, in the absence of loss, for $\omega <  \omega_0$ and $\omega> \omega_\textrm{LO}$ the material is elliptic and the exponents are real. Between $\omega_0$ and $\omega_\textrm{EP}
$ (yellow-shaded region) the material is hyperbolic in the underdamped phase. At the lower Reststrahlen frequency, the exponents transition from real and tending to infinity, to complex conjugate. Beyond the EP, between $\omega_\textrm{EP}$ and $\omega_\textrm{LO}$ (gray-shaded region), the material transitions to the overdamped phase, and the exponents become real.  At the upper limit of the Reststrahlen band, $\omega_\textrm{LO}$, the shell transitions from the overdamped hyperbolic to the elliptic phase. 

Nevertheless, including the optical loss makes the transition not exact, as the condition $\varepsilon_t/\varepsilon_r = -1/[4l(l+1)]$ is not met. Therefore, in the underdamped region the real parts of $t_{1,2
}$ have a small gap and in the overdamped they are not fully real. At the EP frequency, the two solutions do not fully coalesce. However, the signature of the transition is still preserved.

In the overdamped phase, the resonance condition can be met for a isolated frequency, and the shell acts mostly as an attenuator of the electric field. In the underdamped phase, however, due to the field being oscillatory, the boundary conditions can be met for multiple modes, as in a Fabry--P\'erot cavity. Searching for the poles of the coefficient $B_3$ in the lossless limit, the resonance condition takes the form of a phase-matching condition,
\begin{equation}
\Delta \log\frac{a}{b} - \varphi_{BC} = n\pi, \quad n=1,2,3,\ldots
\end{equation}
where  $\varphi_{BC} =
\arg\left\{
(l\varepsilon_c-\varepsilon_r t_1)
[\varepsilon_r t_2+(l+1)\varepsilon_h]
\right\}
$ is a phase that depends on the boundary conditions. This means that the underdamped phase can host a theoretically infinite number of resonances in the absence of losses, as $\Delta$ diverges at $\omega_0$, with the frequencies of higher-order modes clustering at $\omega_0^+$. However, in a realistic picture, the maximum $\Delta$ reached will be finite, and the actual number of modes observed will be limited by optical loss, the thickness of the shell, the 
oscillator strength of the material and the permittivity of the core and host materials.

The phase accumulation is governed by the logarithmic shell thickness, $\log(a/b)$. Thus, for a fixed inner radius, increasing the outer radius increases the phase accumulated across the shell, and each additional $\pi$ of phase adds half an oscillation of the radial profile and a new resonance. This directly links the spatial oscillations induced by the EP transition to the emergence of additional resonances for thicker shells.

In Fig. \ref{fig:Lorentzian}(b) we plot the extinction cross section $\log_{10}({\sigma_\textrm{ext}/(4 \pi a^2)})$, including the radiative correction (see Appendix B) fixing  $b=10$~nm and changing $\log(a/b)$ to study the effect of thickness. As expected, thin shells do not accumulate enough phase to produce multiple resonances, but increasing the thickness, several resonances can be resolved in the extinction. The resonance bands are trapped in the hyperbolic region, between $\omega_0$ and $\omega_\textrm{EP}$. For $a=30$~nm, i.e. $\log(a/b) = \log 3 \approx 1.1$ (dashed horizontal line) the shell hosts up to $n=3$ modes, fixed at $\hbar\omega_n = 2.42$, $2.13$, and $2.05$~eV, marked by white dots in the plot. In Fig. \ref{fig:Lorentzian}(c) we plot the normalized field amplitudes $|\mathbf{E}|/E_{\textrm{max}n}$ at these resonances for $b=10$~nm, with $E_{\textrm{max}n}/E_0 = 29.1$, $13.1$, and $4.2$ for $n=1,2,3$. The progression for $n=1,2,3$ can be seen clearly: each order adds one half oscillation of the radial profile across the shell.

\section{Conclusions}
In this work, we have shown that radial hyperbolicity in a shell
induces an spatial exceptional point and a phase
transition governed by radial anisotropy rather than by gain, loss, or
non-reciprocity. The radial Laplace equation in a uniaxial shell maps to a damped harmonic oscillator, where a logarithmic radius, $s =\log r$ takes the role of time, and the degree of anisotropy acts like an effective stiffness. After reaching a hyperbolicity threshold, the field within the shell switches from monotonic decay (overdamped phase) into spatial oscillation (underdamped phase). The critical point of this transition is an exceptional point, where the two power-law solutions coalesce spatially. The oscillatory field induced by the EP transition to the underdamped phase, enables the emergence of multiple thickness-dependent Fabry--P\'erot-like resonances. This work establishes the connection between spatial exceptional points and the optical response of hyperbolic coreshell particles. 
\section*{Funding}
AB, MSB, and NMRP acknowledge support from the European Union through the EIC PATHFINDER OPEN project No.~101129661-ADAPTATION.

\section*{Data availability}
Data underlying the results presented in this paper are available in \cite{HyperboliEPScode}.

\bibliographystyle{unsrtnat}
\bibliography{hyperbolicEPs}

\appendix
\section{Derivation of electric field and potential coefficients}
In this section we derive the electric field in the anisotropic  core--shell system. The electric potential in a spherical geometry can be found solving the Laplace equation in spherical coordinates\cite{moradi_revisit_2023},  
\begin{equation}
\nabla \cdot(-\hat{\varepsilon}\nabla \phi)=0.
\end{equation}

Choosing the polar axis along the polarization of the applied field, the problem has azimuthal symmetry, $\partial\phi/\partial\varphi = 0$, where $\varphi$ is the azimuthal angle, and the equation can be simplified to
\begin{equation}
\frac{\partial}{\partial r}\left(r^{2} \frac{\partial \phi}{\partial r}\right) + \frac{\varepsilon_t}{\varepsilon_{r}} \frac{1}{\sin \theta} \frac{\partial}{\partial \theta}\left(\sin \theta \frac{\partial \phi}{\partial \theta}\right) = 0.
\end{equation}
By variable separation, $\phi(r,\theta) = R(r)\Theta(\theta)$, we get the differential equation for the radial function, which has the form of a Cauchy-Euler equation,
\begin{equation}
r^2 \frac{d^2 R}{dr^2} + 2r \frac{dR}{dr} - l(l+1) \frac{\varepsilon_t}{\varepsilon_r} R = 0.
\label{eq:radialeq}
\end{equation}
For $\varepsilon_t/\varepsilon_r \neq -1/[4l(l+1)]$, the solutions of this equation are of the form
\begin{equation}
\phi(r,\theta) = \sum_{l=0}^{\infty} \left( A_l r^{t_{1l}} + B_l r^{t_{2l}}\right) P_l(\cos\theta),  \qquad
t_{1l,2l} = \frac{1}{2} \left(-1 \pm \sqrt{1 + 4l(l+1)\frac{\varepsilon_t}{\varepsilon_r}}\right).
\end{equation}

The electric potential in the three regions is, then (the external term $-E_0 r^l P_l(\cos\theta)$ describes a multipolar applied potential; a uniform applied field corresponds to $l=1$)
\begin{equation}
\phi(\textbf{r}) =  \begin{cases}
A_1 r^l P_l(\cos\theta) & r<b, \\[1mm]
(A_2 r^{t_{1l}} + B_2 r^{t_{2l}})P_l(\cos\theta) & b<r<a, \\[1mm]
\left(-E_0 r^l + \frac{B_3}{r^{l+1}}\right) P_l(\cos \theta) & r>a.
\end{cases}
\label{eq:potential}
\end{equation}
The electric field inside and outside the core--shell particle is given by $\textbf{E}(\textbf{r})  = -\nabla\phi(\textbf{r})$, which, by plugging the previous equation, yields (here and below, $P_l'(\cos\theta)\equiv dP_l(\cos\theta)/d\theta$, so that $P_1'(\cos\theta)=-\sin\theta$)
\begin{eqnarray}
\textbf{E}(\textbf{r}) = \begin{cases}
-lA_1r^{l-1}P_l(\cos\theta)\hat{\textbf{r}}
-A_1r^{l-1}P_l'(\cos\theta)\hat{\boldsymbol{\theta}}
& r<b, \\[1mm]
-\left(t_{1l}A_2r^{t_{1l}-1}+t_{2l}B_2r^{t_{2l}-1}\right)
P_l(\cos\theta)\hat{\textbf{r}}\\
\qquad
-\left(A_2r^{t_{1l}-1}+B_2r^{t_{2l}-1}\right)
P_l'(\cos\theta)\hat{\boldsymbol{\theta}}
& b<r<a, \\[1mm]
\left[lE_0r^{l-1}+(l+1)\frac{B_3}{r^{l+2}}\right]
P_l(\cos\theta)\hat{\textbf{r}}\\
\qquad
+\left[E_0r^{l-1}-\frac{B_3}{r^{l+2}}\right]
P_l'(\cos\theta)\hat{\boldsymbol{\theta}}
& r>a.
\end{cases}
\label{eq:electric_field}
\end{eqnarray}
In order to obtain the expressions for the unknown coefficients, we apply the usual boundary conditions at the two interfaces, i.e. the continuity of the radial component of the electric displacement, $D_r =  \varepsilon_r E_r$, and the continuity of the tangential component of the electric field, $E_\theta$, which read

\begin{equation}
\begin{cases}
l\varepsilon_c A_1b^{l-1}
=\varepsilon_r\left(t_{1l}A_2b^{t_{1l}-1}
+t_{2l}B_2b^{t_{2l}-1}\right), \\[1mm]
A_1b^{l-1}
=A_2b^{t_{1l}-1}+B_2b^{t_{2l}-1},\\[1mm]
-\varepsilon_r\left(t_{1l}A_2a^{t_{1l}-1}
+t_{2l}B_2a^{t_{2l}-1}\right)
=\varepsilon_h\left[
lE_0a^{l-1}+(l+1)\frac{B_3}{a^{l+2}}
\right],\\[1mm]
-\left(A_2a^{t_{1l}-1}+B_2a^{t_{2l}-1}\right)
=E_0a^{l-1}-\frac{B_3}{a^{l+2}}.
\end{cases}
\end{equation}
Solving the system of equations, we find the following amplitudes for the electric potential,
\begin{align}
A_1 ={}&
-\frac{
(2l+1)\varepsilon_r\varepsilon_h(t_{2l}-t_{1l})
E_0a^{l-1}b^{1-l}
}{
\begin{aligned}
&(\varepsilon_rt_{2l}-l\varepsilon_c)
(\varepsilon_rt_{1l}+(l+1)\varepsilon_h)
\left(\frac{b}{a}\right)^{1-t_{1l}}
-&(\varepsilon_rt_{1l}-l\varepsilon_c)
(\varepsilon_rt_{2l}+(l+1)\varepsilon_h)
\left(\frac{b}{a}\right)^{1-t_{2l}}
\end{aligned}
}, \\
A_2 ={}&
-\frac{
(2l+1)\varepsilon_h(\varepsilon_rt_{2l}-l\varepsilon_c)
E_0a^{l-1}b^{1-t_{1l}}
}{
\begin{aligned}
&(\varepsilon_rt_{2l}-l\varepsilon_c)
(\varepsilon_rt_{1l}+(l+1)\varepsilon_h)
\left(\frac{b}{a}\right)^{1-t_{1l}}
-&(\varepsilon_rt_{1l}-l\varepsilon_c)
(\varepsilon_rt_{2l}+(l+1)\varepsilon_h)
\left(\frac{b}{a}\right)^{1-t_{2l}}
\end{aligned}
}, \\
B_2 ={}&
\frac{
(2l+1)\varepsilon_h(\varepsilon_rt_{1l}-l\varepsilon_c)
E_0a^{l-1}b^{1-t_{2l}}
}{
\begin{aligned}
&(\varepsilon_rt_{2l}-l\varepsilon_c)
(\varepsilon_rt_{1l}+(l+1)\varepsilon_h)
\left(\frac{b}{a}\right)^{1-t_{1l}}
-&(\varepsilon_rt_{1l}-l\varepsilon_c)
(\varepsilon_rt_{2l}+(l+1)\varepsilon_h)
\left(\frac{b}{a}\right)^{1-t_{2l}}
\end{aligned}
}, \\
B_3 ={}&
\frac{
E_0a^{2l+1}
\left[
(\varepsilon_rt_{2l}-l\varepsilon_c)
(\varepsilon_rt_{1l}-l\varepsilon_h)
\left(\frac{b}{a}\right)^{1-t_{1l}}
-
(\varepsilon_rt_{1l}-l\varepsilon_c)
(\varepsilon_rt_{2l}-l\varepsilon_h)
\left(\frac{b}{a}\right)^{1-t_{2l}}
\right]
}{
\begin{aligned}
&(\varepsilon_rt_{2l}-l\varepsilon_c)
(\varepsilon_rt_{1l}+(l+1)\varepsilon_h)
\left(\frac{b}{a}\right)^{1-t_{1l}}
-&(\varepsilon_rt_{1l}-l\varepsilon_c)
(\varepsilon_rt_{2l}+(l+1)\varepsilon_h)
\left(\frac{b}{a}\right)^{1-t_{2l}}
\end{aligned}
}.
\end{align}

The previous solutions are only valid as long as $\varepsilon_t/\varepsilon_r \neq -1/[4l(l+1)]$. For the special case of the exceptional point, $\varepsilon_t/\varepsilon_r = -1/[4l(l+1)]$, the characteristic equation has a repeated root, $t=-1/2$. The two linearly independent radial solutions are then $r^{-1/2}$ and $r^{-1/2}\log(r/b)$. The general solution for the potential in the case of repeated roots is \cite{trench2013}
\begin{equation}
\phi(r,\theta) =
r^{-\frac{1}{2}}\left( A_2 + B_2 \log \frac{r}{b} \right)P_l(\cos\theta).
\end{equation}
Within this special case of the exceptional point, the electric potential in the three regions is, then
\begin{equation}
\phi(\textbf{r}) =  \begin{cases}
A_1 r^l P_l(\cos\theta) & r<b, \\[1mm]
r^{-\frac{1}{2}}\left( A_2 + B_2 \log \frac{r}{b} \right)P_l(\cos\theta)
& b<r<a, \\[1mm]
\left(-E_0 r^l + \frac{B_3}{r^{l+1}}\right)P_l(\cos \theta)
& r>a,
\end{cases}
\label{eq:potential2}
\end{equation}
and the electric field inside and outside the core--shell particle is
\begin{eqnarray}
\textbf{E}(\textbf{r}) = \begin{cases}
-lA_1r^{l-1}P_l(\cos\theta)\hat{\textbf{r}}
-A_1r^{l-1}P_l'(\cos\theta)
\hat{\boldsymbol{\theta}}
& r<b, \\[1mm]
r^{-\frac{3}{2}}
\left[
\frac{1}{2}A_2-B_2
+\frac{1}{2}B_2\log \frac{r}{b}
\right]P_l(\cos\theta)\hat{\textbf{r}}\\
\qquad
-r^{-\frac{3}{2}}\left(A_2+B_2\log \frac{r}{b}\right)
P_l'(\cos\theta)
\hat{\boldsymbol{\theta}}
& b<r<a, \\[1mm]
\left[lE_0r^{l-1}+(l+1)\frac{B_3}{r^{l+2}}\right]
P_l(\cos\theta)\hat{\textbf{r}}\\
\qquad
+\left[E_0r^{l-1}-\frac{B_3}{r^{l+2}}\right]
P_l'(\cos\theta)
\hat{\boldsymbol{\theta}}
& r>a.
\end{cases}
\label{eq:electric_field2}
\end{eqnarray}
where the coefficients are given by
\begin{align}
A_1 &=
-\frac{
4(2l+1)a^{l+\frac12}b^{-l-\frac12}
E_0\varepsilon_h\varepsilon_r
}{
4\varepsilon_r\left[l\varepsilon_c+(l+1)\varepsilon_h\right]
+\left[
4l(l+1)\varepsilon_c\varepsilon_h
+2\varepsilon_r\left((l+1)\varepsilon_h-l\varepsilon_c\right)
-\varepsilon_r^2
\right]
\log\left(\frac{a}{b}\right)
},
\\[1ex]
A_2 &=
-\frac{
4(2l+1)a^{l+\frac12}E_0
\varepsilon_h\varepsilon_r
}{
4\varepsilon_r\left[l\varepsilon_c+(l+1)\varepsilon_h\right]
+\left[
4l(l+1)\varepsilon_c\varepsilon_h
+2\varepsilon_r\left((l+1)\varepsilon_h-l\varepsilon_c\right)
-\varepsilon_r^2
\right]
\log\left(\frac{a}{b}\right)
},
\\[1ex]
B_2 &=
-\frac{
2(2l+1)a^{l+\frac12}E_0\varepsilon_h
(2l\varepsilon_c+\varepsilon_r)
}{
4\varepsilon_r\left[l\varepsilon_c+(l+1)\varepsilon_h\right]
+\left[
4l(l+1)\varepsilon_c\varepsilon_h
+2\varepsilon_r\left((l+1)\varepsilon_h-l\varepsilon_c\right)
-\varepsilon_r^2
\right]
\log\left(\frac{a}{b}\right)
},
\\[1ex]
B_3 &=
\frac{
a^{2l+1}E_0
\left\{
4l\varepsilon_r(\varepsilon_c-\varepsilon_h)
-\left[
4l^2\varepsilon_c\varepsilon_h
+2l\varepsilon_r(\varepsilon_c+\varepsilon_h)
+\varepsilon_r^2
\right]\log\left(\frac{a}{b}\right)
\right\}
}{
4\varepsilon_r\left[l\varepsilon_c+(l+1)\varepsilon_h\right]
+\left[
4l(l+1)\varepsilon_c\varepsilon_h
+2\varepsilon_r\left((l+1)\varepsilon_h-l\varepsilon_c\right)
-\varepsilon_r^2
\right]\log\left(\frac{a}{b}\right)
}.
\end{align}
\section{Polarizability and extinction cross section}

The quasistatic polarizability is obtained directly from the coefficient \(B_3\) of the induced dipole term in the external potential,
\begin{equation}
\alpha_{\mathrm{qs}}(\omega)
=4\pi\varepsilon_0\varepsilon_h\frac{B_3}{E_0}.
\end{equation}

To account for radiative losses, we introduce the radiative correction to the quasistatic polarizability:

\begin{equation}
\alpha_{\mathrm{rad}}(\omega)=
\frac{\alpha_{\mathrm{qs}}(\omega)}
{1-i\displaystyle\frac{k^3}{6\pi\varepsilon_0\varepsilon_h}
\alpha_{\mathrm{qs}}(\omega)},
\end{equation}
where $k=\sqrt{\varepsilon_h}\,\omega/c$ is the wavenumber in the host medium. Finally, the extinction cross section is obtained from the dipolar optical theorem:

\begin{equation}
\sigma_{\mathrm{ext}}(\omega)
=\frac{k}{\varepsilon_0\varepsilon_h}
\operatorname{Im}\left[\alpha_{\mathrm{rad}}(\omega)\right].
\end{equation}

\end{document}